\documentclass[prl,aps,floatfix,superscriptaddress,nofootinbib,twocolumn,showpacs]{revtex4}
\usepackage{graphicx}
\usepackage{setspace}
\usepackage{epstopdf}
\usepackage{subfigure}
\usepackage{amsmath}
\usepackage{amssymb}
\usepackage{amsthm}
\usepackage{amscd}
\usepackage[usenames]{color}

\begin{document}

\title{Threading the Needle:  Generating Textures in Nematics}
\author{Bryan Gin-ge Chen}
\affiliation{Instituut-Lorentz, Universiteit Leiden, Postbus 9506, 2300 RA Leiden, The Netherlands}
\affiliation{Department of Physics \& Astronomy, University of Pennsylvania, Philadelphia PA 19104, USA}
\author{Paul J. Ackerman}
\affiliation{Department of Physics, University of Colorado, Boulder, CO 80309, USA}
\author{Gareth P. Alexander}
\affiliation{Department of Physics and Centre for Complexity Science, University of Warwick, Coventry, CV4 7AL, UK}
\author{Randall D. Kamien}
\affiliation{Department of Physics \& Astronomy, University of Pennsylvania, Philadelphia PA 19104, USA}
\author{Ivan I. Smalyukh}
\affiliation{Department of Physics, University of Colorado, Boulder, CO 80309, USA}
\date{\today}

\pacs{61.30.Jf, 61.30.Dk, 11.10.Lm}
\begin{abstract}
The Hopf fibration is an example of a texture: a topologically stable, smooth, global configuration of
a field.  Here we demonstrate the controlled sculpting of the Hopf
fibration in nematic liquid crystals through the control of point
defects.  We demonstrate how these are related to torons by use of a
topological visualization technique derived from the Pontryagin-Thom
construction.
\end{abstract}
\maketitle

The combination of geometric order, optical response, and soft
elasticity of liquid crystals uniquely positions them as an arena to
study topology: boundary conditions on sample walls can obstruct
smooth solutions in the bulk, forcing points, lines, and walls of
diminished order \cite{Klemanbook}.  In general, these topological
defects serve as tools for probing the symmetries of the ground state
manifold (GSM) \cite{Mermin}, as fundamental excitations \cite{KT},
and as potential building blocks for self-assembly \cite{Nelsontetra}.
Does topology only play a role in systems with {\sl singularities}?
Certainly not, the Skyrmion in two and three dimensions is an
everywhere smooth complexion of order that is, nonetheless,
topologically protected \cite{Skyrme,MantonSutcliffe}.  Similar non-singular configurations are the origin, for instance, of gapless excitations in the quantum hall effect and topological insulators \cite{TKNN,KaneandMele}.  In the case of
nematic liquid crystals, the GSM is $\mathbb{R}P^2$ and two- and
three- dimensional Skyrmions are labelled by elements of the second
and third homotopy groups, $\pi_2(\mathbb{R}P^2)$ and
$\pi_3(\mathbb{R}P^2)$, respectively. The non-trivial elements of the
latter correspond to the much-storied ``Hopf fibration'' \cite{Hopf}, an
allowed texture in the nematic phase \cite{bouligand,BDPPT}.  In this
Letter, we demonstrate our ability to controllably generate and robustly visualize the Hopf
fibration in cholesteric systems: nematics with a
preferential handedness, or twist.

Our starting point is the toron configuration depicted in Fig.~1 and
described in detail in \cite{toron}.  This is a tube of double-twist
 that is wrapped upon itself, its boundary
forming a torus.  Above and below the ``donut hole,'' there are two
point defects, both taking the form of hyperbolic hedgehogs.  By
manipulating these two point defects we can create a defect free
texture with the topology of the Hopf fibration as in Fig. \ref{fig:hopf}. Recall that the preimage of an element of the GSM,  ${\bf\hat n}_0\in\mathbb{R}P^2$, is the set of points in the sample where the director
field ${\bf\hat n}={\bf\hat n}_0$.  Because we are considering three-dimensional nematics, we have a map from a three dimensional space (the sample) to the two-dimensional GSM: it follows that the preimage of ${\bf\hat n}_0$ is a
closed curve with topology of one or more disjoint circles. The Hopf fibration is characterized by the linking of the preimages of {\sl any} two directions ${\bf\hat n}_1\ne{\bf\hat n}_2$.  

How do we know it is the Hopf fibration, and how does the topology
work out to render this result?  To answer the first question, we
utilize the Pontryagin-Thom construction for visualization of nematic
configurations \cite{tD,BGC} which we now briefly sketch.  
This method is a three-dimensional
generalization of the use of crossed polarizers to study Schlieren
textures in two-dimensional samples with the director taking values in
$\mathbb{R}P^1$ \cite{TeoKane}.

Recall
that the dark lines in a Schlieren texture mark those regions where
the director is along one of the two polarizer directions; continuity
ensures that the dark lines only end on point defects and topology
ensures that an even number of dark lines emanate from each point
defect.  
We can abstract this slightly by considering only half the lines,
{\sl i.e.}, those corresponding to just one of the polarizer directions; note
that in either case, each line carries an arbitrarily chosen yet {\sl globally} consistent orientation so that the
lines point, say, from positive to negative defects.
\begin{figure*}
\includegraphics[width=0.55\textwidth]{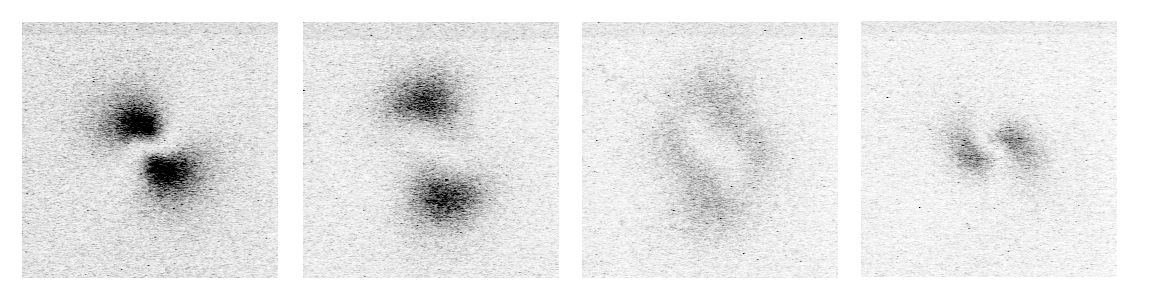}
\\
\includegraphics[width=0.35\textwidth]{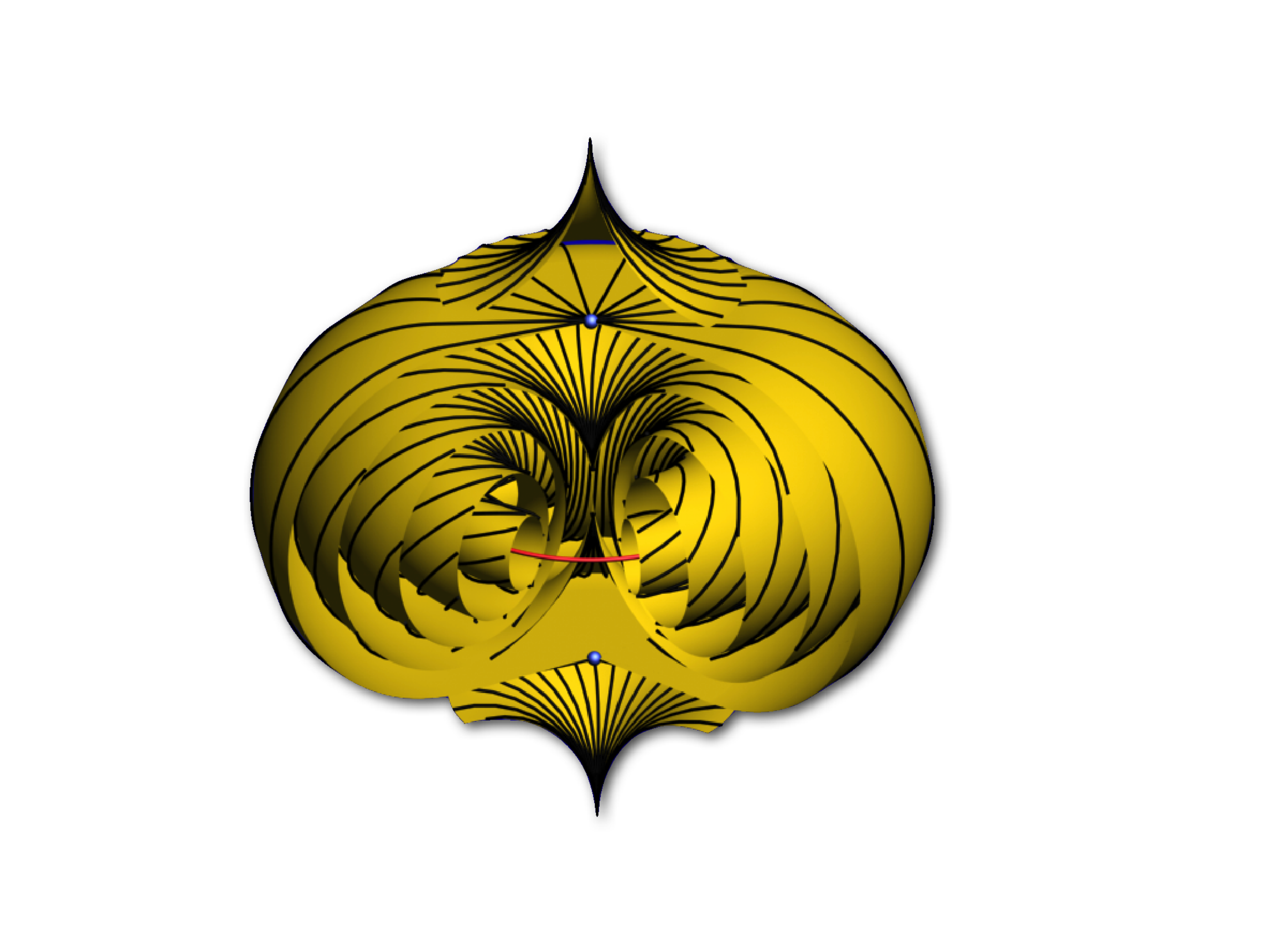}\includegraphics[width=0.65\textwidth]{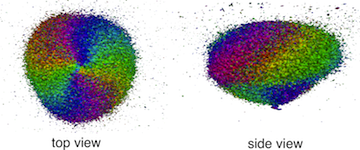}
\caption{
The top figures shows selected 3PEF-PM images from an image stack -- the
images are 16 $\mu$m wide and approximately 4 $\mu$m apart in the
$z$-direction.  Bottom Left: the toron texture from~\cite{toron}.
Bottom Center and Right: the ``Pontryagin-Thom'' surface constructed
from this image stack.  The spots of color are an artifact of our
construction scheme -- they are places where the software has chosen
the wrong branch.  Fortunately, the robustness of this method allows
us to recognize the overall texture.  Note the two hedgehog defects on top and bottom.}
\label{fig:toron}
\end{figure*}

This two-dimensional construction has a natural generalization to
three dimensions.  We first pick a probe direction ${\bf\hat
p}\in\mathbb{R}P^2$.  Next we draw the surface, $\Sigma_{\bf\hat p}\in\mathbb{R}^3$, on which the
director is everywhere {\sl perpendicular} to the probe, ${\bf\hat
p}\cdot {\bf\hat n}=0$.  We are therefore looking at the preimage of a whole
curve $\bf\hat p_\perp$ in $\mathbb{R}P^2$, the ``equator'' if ${\bf\hat p}$ were the ``North pole.''  Were we to look for the preimage of $\bf\hat
p$ we would generically only get a curve and, more problematically, we
could get the empty set for a nontrivial texture.  The surface construction, however, neatly
generalizes the two-dimensional case: the boundaries of any surface
must be topological defects.  A line boundary is the location of a
disclination line, carrying the $\mathbb{Z}_2$ charge associated with
$\pi_1(\mathbb{R}P^2)$, while a point boundary, {\sl i.e.} a hole in
the surface, carries a $\mathbb{Z}$ charge associated with
$\pi_2(\mathbb{R}P^2)$ where here and throughout we only use based homotopy groups.  The surfaces, however, do not carry enough
information to determine the point charges.  In order to capture this
information, we must add an additional piece of information to the
surface, namely the direction of the director in $\Sigma_{\bf\hat p}$.
We represent this pictorially through a color wheel, ranging from red
to violet through orange, yellow, green, blue, and indigo.
as the director rotates by $\pi$.  We pass through the color wheel a
second time if the director rotates by $2\pi$, as it does, for
instance, in $\Sigma_{\bf\hat z}$ for the standard radial hedgehog
shown in Fig.~\ref{fig:PT}.  In fact, since all point defects in a uniaxial nematic can be
oriented \cite{RMP}, the usual ${\bf\hat n}\rightarrow -{\bf\hat n}$ symmetry
does not come into play so the director will always rotate through the
color wheel an even number of times, that is, rotations by multiples
of $2\pi$. As a result, a point defect of charge
$p\in\pi_2(\mathbb{R}P^2)$ will have a winding of $2p\pi$, or will
cover the color wheel $2p$ times, providing a unique identification of
point defects in nematics.   

\begin{figure*}
\includegraphics[width=0.27\textwidth]{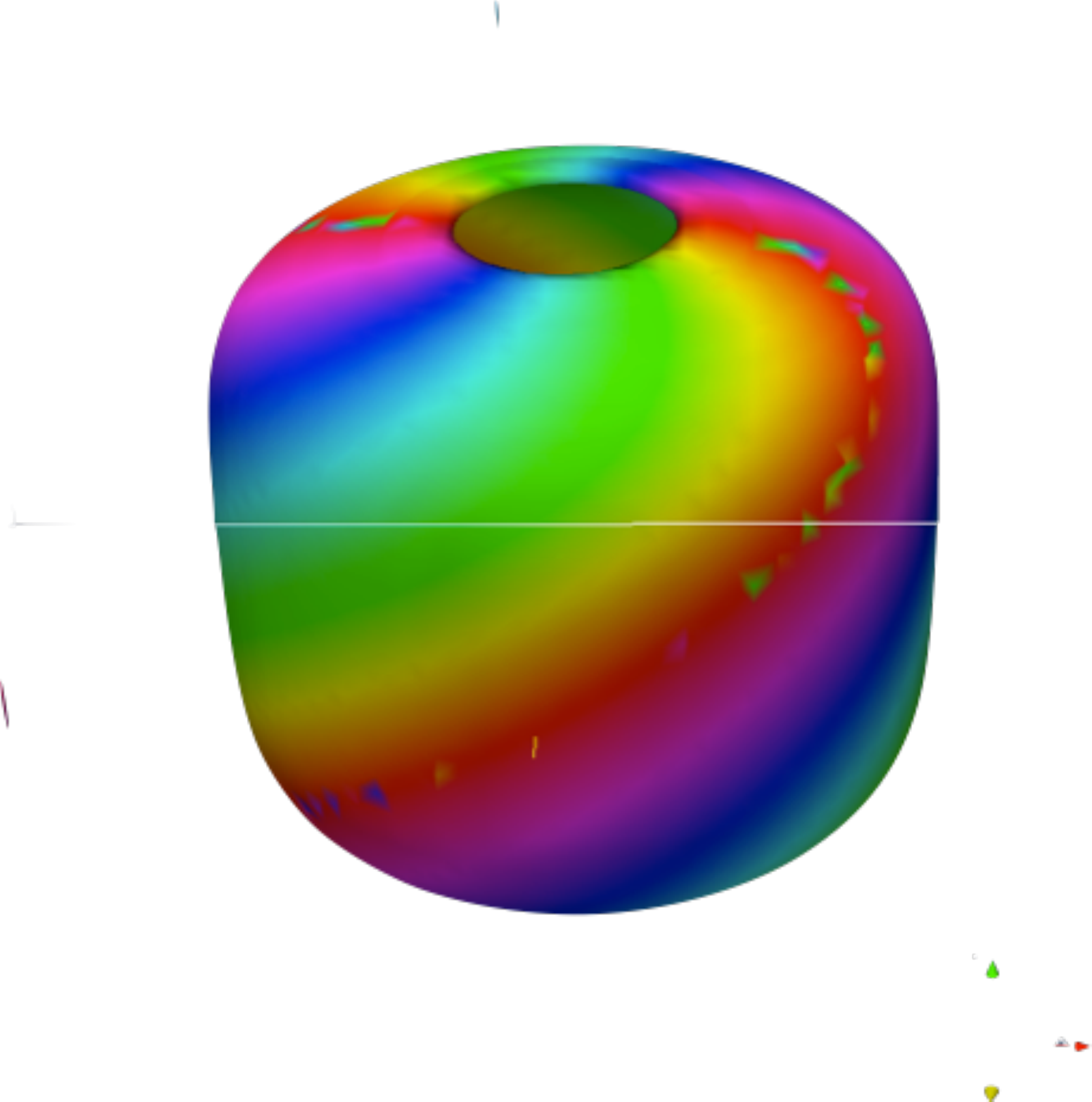}\hskip0.45truein\includegraphics[width=0.54\textwidth]{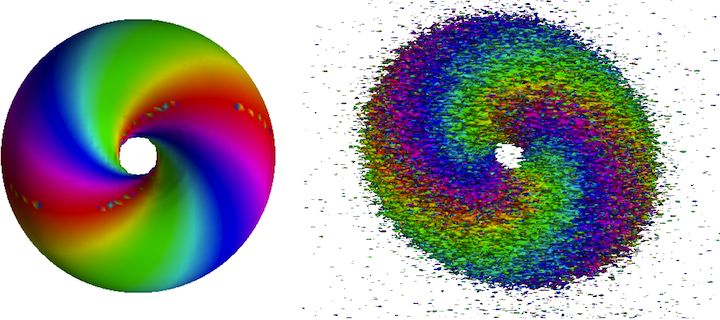}
\caption{On the left we show a simulation texture of
a toron in which the point hedgehogs are replaced with disclination
loops.  Bringing those loops together through the center of the spool
gives a torus with linked preimages.  In the center, we show a
simulation of the resulting Hopf fibration.  On the right, we show an
experimental image of the same texture.}\label{fig:hopf}
\end{figure*}

There is an important constraint on the colors that may paint the
preimage.  The neighborhood of the preimage $\Sigma_{\bf\hat p}$ must admit a
continuous map to the GSM \cite{BGC} and the neighborhood of ${\bf\hat p}_\perp\in\mathbb{R}P^2$ is a
M\"obius strip M, a line bundle over the equatorial circle. This means
that the normal vectors to points in $\Sigma_{\bf\hat p}$ are mapped continuously to
points of M lying``above'' the points of the base circle.  Since one
turn through the color wheel is half a trip through M , the surface
normal on any path in $\Sigma_{\bf\hat p}$ whose image in $\mathbb{R}P^2$ wraps the equator once
will reverse sign Ð this extra structure forces every closed curve on
$\Sigma_{\bf\hat p}$ to have an even color winding so that the image of the surface
normal in M can be continuous.

Finally, it can be shown that this representation is faithful, that is, up to
homotopy, no
information is lost and the original texture can be constructed from the 
representation we present here \cite{tD}. A further
advantage is that complicated homotopies of three-dimensional
configurations can be visualized by manipulating the surfaces; not
only continuous deformations but also a general class of merging of
surfaces called bordisms.

In the experiments, we used nematic LC ZLI 2806 doped with the chiral agent CB15 to obtain the
cholesteric pitch $P$ of interest according to the relationship
concentration of CB15 $C= 1/ (h * P)$, where $h=5.9$
$\mu\text{m}^{-1}$ is the helical twisting power of the used
combination of the nematic host and chiral additive. Our cholesteric mixture had
$P=20\mu$m to match the thickness of the used capillary $d$ so that $d/P=1$. A
rectangular capillary with $20 \times 200\mu$m cross-section was treated for
vertical surface boundary conditions by infiltrating it with a 0.1wt.\%
aqueous solution of a surfactant
[3-(trimethoxysilyl)propyl]octadecyldimethylammonium chloride (DMOAP)
and then evaporating the solution by heating it to 90$^{\circ}$C and keeping it
at this temperature for about 30 min. The cholesteric mixture was then
heated to isotropic phase at 80$^{\circ}$C and infiltrated to the capillary to
avoid filling-induced defects. Various twist-stabilized localized
structures in an initially unwound frustrated cholesteric LC were
formed through the use of holographic optical tweezers (HOT) \cite{1} built
around a spatial light modulator (SLM) and a CW laser operating at
1064nm. Laser beams of power less than 50mW were focused and spatially
steered in 3D within the sample. We have used 10X-100X microscope
objectives with numerical apertures ranging within NA=0.1-1.4 for
optical generation.

Imaging of the samples utilized 
three-photon excitation
fluorescence polarizing microscopy (3PEF-PM) \cite{2} integrated with HOT
into a single optical setup built around the same inverted optical
microscope IX-81 (Olympus). 
The optical technique of 3PEF-PM \cite{2} is
non-invasive, does not require dyes (since the detected fluorescence
comes from the LC molecules themeselves), and enables the imaging of
director fields in 3D.  
The non-linear three-photon absorption process
gives rise to a $\cos^6\beta$ orientational dependence of the fluorescence
signal, where $\beta$ is the angle between the probing light's linear
polarization and the director. The inherent z-resolution (along the
microscope's optical axis) associated with the non-linear process
allows for optical sectioning and reconstruction of 3D images of the
director field.  3D 3PEF-PM images for four linear polarizations are
used to generate a representation in Paraview \cite{paraview}.

Whereas Schlieren textures in thin cells give directly the
Pontryagin-Thom construction for (quasi-)two-dimensional nematics, the
analogous colored surfaces of three-dimensional textures are not an
automatic output of any current imaging technique. These surfaces can
be extracted easily from knowledge of the director field, which can in
turn be obtained from confocal microscopy \cite{smalyukh01,smalyukh02} or polarizing-mode nonlinear
optical microscopies such as 3PEF-PM, coherent anti-Stokes Raman
scattering microscopy \cite{AA}, and stimulated Raman scattering
microscopy \cite{BB}. 
To construct this surface, we take intensity data from confocal
slices, polarized at four different angles $\pi/4$ apart in the $xy$
plane $(E_0^2,E_{\pi/4}^2,E_{\pi/2}^2,E_{3\pi/4}^2)$.  The Stokes 
parameters $I,Q$, and $U$ are:
\begin{align}\label{eq:stokes}
I&=\frac{1}{2}\left(E_0^2+E_{\pi/4}^2+E_{\pi/2}^2+E_{3\pi/4}^2\right),\\
Q&=E_0^2-E_{\pi/2}^2,\\
U&=E_{\pi/4}^2-E_{3\pi/4}^2.
\end{align}
Writing ${\bf\hat
n}=[\sin\theta\cos\phi,\sin\theta\sin\phi,\cos\theta]^T$ and taking
the electric field amplitude to simply be proportional to the local electric anisotropy tensor we find that $I\propto J\sin^n\theta$
and $Q/U=\tan(2\phi)$ where $J$ is the amplitude of the signal. Here
$n$ is an exponent depending on the imaging modality; $n=4$ for the
case of
fluorescence confocal microscopy \cite{smalyukh01,smalyukh02}, $n=6$ for 3PEF-PM with fluorescence
detection without a polarizer \cite{2}, and $n=8$ for coherent
anti-Stokes Raman scattering polarizing microscopy with linearly
polarized detection collinear with the polarization of excitation
light. Away from the toron the director is normal to the top and bottom surface and so $\theta=0$ there.  Thus we may shift and 
normalize the calculated $I$ from
the data to take values from $0$ to $1$ so that the $n$th
root of $I$ gives us $\sin\theta$.  The angle $\phi$ then gives us the angle of
the polarization projected to the $xy$ plane, and we can reconstruct
the director ${\bf\hat n}$ from $\theta$ and $\phi$.

To go from this to the colored surface numerically, we reflect the
director field so that it lies in the upper half of the sphere, {\sl
i.e.} if $\cos\theta<0$ we take ${\bf\hat n}\rightarrow -{\bf\hat n}$.
Using ParaView \cite{paraview} we then view the isocontour with $n_z$
close to zero. Though one might want to take a slice with $n_z$ zero,
the non-orientability of the line field makes it difficult to exclude
the artificial ``branch cuts'' where any reconstruction assigns
adjacent grid points to different branches of ${\bf\hat n}$, for
example when ${\bf\hat n}$ happens to be adjacent to a data point of
$-{\bf\hat n}$.  The downside of our approach is that what should be
one surface at $n_z=0$ is actually two nearby surfaces
$n_z=\pm\epsilon$.  Note that all we pick out here are the surfaces of
(near) maximum $I$, so the sixth-root transformation we made above actually makes no difference; all we need is the fact that the regions in the data where $I$ is maximum correspond to regions where the molecules tend to lie in the $xy$ plane. 
\begin{figure*}
\includegraphics[width=0.9\textwidth]{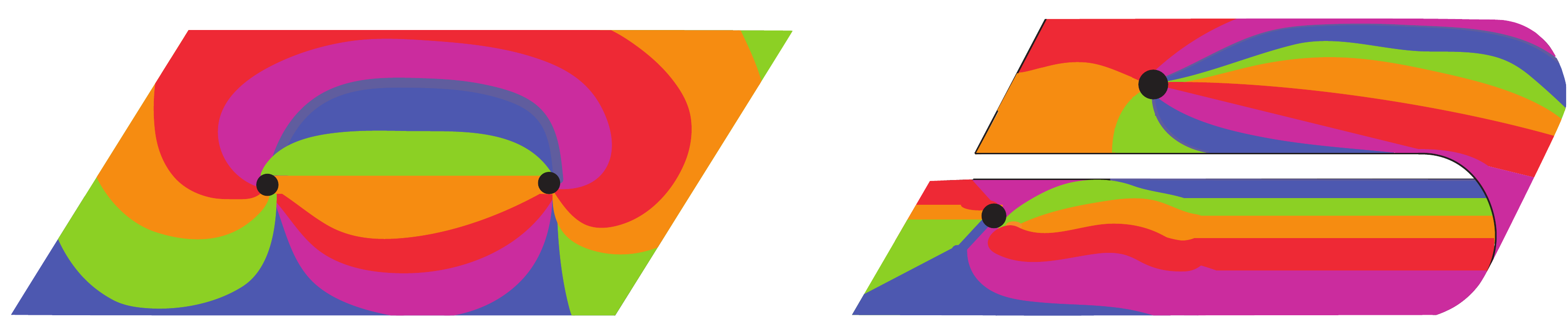}
\caption{\label{fig:PT}On the left we show a schematic of a
hedgehog/anti-hedgehog pair.  They are of opposite sign because the
sense of their color rotations is of opposite handedness.  
Note, however, that viewed from below the handedness of each color wheel
changes, illustrating the global sign ambiguity.  On the right we show
a schematic of the same pair of defects moved one on top of each other
with the surface bent around.  Were we to neglect the orientation
imposed by the surface normal, we might think the two defects are of
the same sign.  The surface keeps track of the global base point.}
\end{figure*}

We analyzed 3PEF-PM images of several chirally doped nematic textures as described.  A precise reconstruction of the director field requires a careful
analysis of the optical properties of the material. However, this is
unnecessary to determine topological features, which are independent
of the fine details and depend only on the coarse structure that is
preserved under continuous deformation. Thus even a highly approximate
reconstruction of the director will capture all of the topology
correctly.

Using these tools, we show the toron texture in this Pontryagin-Thom representation in Fig.~\ref{fig:toron}, alongside the director picture \cite{toron}.  The colored stripes on the squashed sphere depict the double-twist inside the torus.  On top and bottom the color twists around signaling a hedgehog/anti-hedgehog pair.  The surface has no boundary and thus there are no disclination lines.  It follows that we can globally orient the texture (up to an overall sign choice) to make it a vector on $S^2$.  Thus, by construction, the surface separates space into an outer region where the director lifts to a vector in the northern hemisphere $n_z>0$ and an inner region where it maps to the southern hemisphere $n_z<0$.  This immediately tells us that there is {\sl not} a Hopf fibration: recall that to measure the degree of the Hopf fibration one takes the preimage of two directions on $S^2$, each a closed loop, and calculates the linking number of the two preimages.  The linking number is independent of which two preimages we take and so, if we take a loop where the vector field is in the southern hemisphere it {\sl can not} link with a preimage of a vector in the northern hemisphere -- the two loops are separated from each other by the surface in Fig.~\ref{fig:toron}!  To create a non-zero linking number, we must somehow entangle the preimages.  Fortunately, the nematic symmetry allows us to easily visualize this.

In nematics, disclination loops can carry hedgehog charge \cite{RMP}.
As seen in experiment \cite{toron} the two point defects in
Fig.~\ref{fig:toron} can both open up into disclination loops.  Now
there is a passageway between the inside and the outside of the
Pontryagin-Thom surface and we can no longer orient the director
field.  Nonetheless, we can now bring the two loops together through
the eye of the spool shown on the left in Fig.~\ref{fig:hopf}.  Once
brought together they cancel and we again have a situation with no
disclinations.  Again, we can orient the director field and the torus
separates the northern and southern hemispheres.  We now see in both
the simulated textures and, more importantly, in the experiment that the
constant colored circles on the torus link with each other -- the
preimages of different directions are linked!  We have created a
degree-one Hopf fibration starting from the toron.  

In closing, we
note that this graphical representation immediately makes clear a number of
often subtle issues in the description of defects in nematics
~\cite{RMP}.  First, we can see how the relative charges of two defects
depends upon a base point: in this representation, a positive point
defect will have a counter-clockwise-rotating color wheel, while a
negative point defect will have a clockwise-rotatating color wheel
when we look from above.  Were we to look at the same surface from
below, however, the handedness of the rotations flip!  This
corresponds to the global ambiguity in choosing charge associated with
the two choices of lifting $\mathbb{R}P^2$ to $S^2$.  It follows that
looking at two pieces of surface in the vicinity of two defects does
not allow the calculation of their relative degree -- one surface must
be used in order to consistently determine the topological charge.
Finally note that these surfaces can end on disclination lines, just
as the dark brushes in the Schlieren texture can end on disclination
points in two-dimensions. Importantly, the construction of a colored
surface from any given liquid crystal texture captures all of the
topological information about the texture, and also permits the full
director field to be reconstructed, at least up to homotopy. 
In future work we will use this method to visualize blue phases and other complex textures.  Generalizing to biaxial nematics is another extension worth pursuing.

We acknowledge stimulating discussions with D. Beller, F. Cohen, and R. Kusner.
GPA, BGC, and RDK were supported in part by NSF DMR05-47230 and a gift
from L.J. Bernstein.  This research was supported in part by the
National Science Foundation under Grant No. NSF PHY11-25915.  GPA,
BGC, RDK, and IIS thank the KITP for their hospitality while this work
was being prepared.  BGC thanks the hospitality of the Boulder School
in Condensed Matter and Materials Physics where some of this work was
completed.

\end{document}